\documentclass[sigconf]{acmart}

\usepackage{booktabs}
\usepackage[normalem]{ulem}
\useunder{\uline}{\ul}{}
\usepackage{amsmath}
\usepackage[table]{}
\usepackage{multirow}
\usepackage{colortbl}
\usepackage{bbm}
\usepackage{bbding}
\usepackage{pifont}
\usepackage{hyperref}
\hypersetup{
    colorlinks=true,
    linkcolor=blue,
    filecolor=magenta,      
    urlcolor=blue,
    pdftitle={Overleaf Example},
    pdfpagemode=FullScreen,
    }
\newcommand{\xmark}{\text{\ding{55}}}
\author{Hansi Zeng}
\affiliation{
  \institution{University of Massachusetts Amherst}
  \country{United States}
}
\email{hzeng@cs.umass.edu}

\author{Hamed Zamani}
\affiliation{
  \institution{University of Massachusetts Amherst}
  \country{United States}
}
\email{zamani@cs.umass.edu}

\author{Vishwa Vinay}
\affiliation{
  \institution{Adobe Research}
  \country{India}
}
\email{vinay@adobe.com}

\usepackage{xspace}
\newcommand{\framework}{CL-DRD\xspace}

\usepackage{algorithm}
\usepackage{algpseudocode}

\acmConference[SIGIR '22]{Make sure to enter the correct
  conference title from your rights confirmation emai}{June 03--05,
  2022}{Madrid, Spain}
  
\copyrightyear{2022}
\acmYear{2022}
\setcopyright{acmcopyright}\acmConference[SIGIR '22]{Proceedings of the 45th International ACM SIGIR Conference on Research and Development in Information Retrieval}{July 11--15, 2022}{Madrid, Spain}
\acmBooktitle{Proceedings of the 45th International ACM SIGIR Conference on Research and Development in Information Retrieval (SIGIR '22), July 11--15, 2022, Madrid, Spain}
\acmPrice{15.00}
\acmDOI{10.1145/3477495.3531791}
\acmISBN{978-1-4503-8732-3/22/07}
  
\begin{document}

%\title{Teaching a Dense Retriever with Curriculum Learning}
%\title{A Curriculum Learning Approach for Optimizing Dense Retrieval using Knowledge Distillation}
\title{Curriculum Learning for Dense Retrieval Distillation}

\begin{abstract}
Recent work has shown that more effective dense retrieval models can be obtained by distilling ranking knowledge from an existing base re-ranking model. In this paper, we propose a generic \textit{curriculum learning} based optimization framework called \framework that controls the difficulty level of training data produced by the re-ranking (teacher) model. \framework iteratively optimizes the dense retrieval (student) model by increasing the difficulty of the knowledge distillation data made available to it. In more detail, we initially provide the student model coarse-grained preference pairs between documents in the teacher's ranking, and progressively move towards finer-grained pairwise document ordering requirements. In our experiments, we apply a simple implementation of the \framework framework to enhance two state-of-the-art dense retrieval models. Experiments on three public passage retrieval datasets demonstrate the effectiveness of our proposed framework.  
\end{abstract}



\maketitle

%\vspace{.05cm}

\section{Introduction}

Retrieval that combines high dimensional vector representations of queries and documents obtained from deep neural networks and approximate nearest neighbor search algorithms have recently attracted considerable attention~\cite{Karpukhin2020DensePR,Xiong2021ApproximateNN,Prakash:2021:RANCE}. These dense retrieval models rely on the availability of large-scale training data, which includes public datasets such as MS MARCO~\cite{Campos2016MSMA}, and proprietary datasets collected from the query logs of deployed search engines. While the numbers of queries and documents are quite large, the datasets often suffer from incomplete relevance judgments \cite{Prakash:2021:RANCE}, i.e., very few documents are judged for a given query. An approach to address this sparsity issue is to train dense retrieval models using \textit{knowledge distillation}. Recent work \cite{Lin2020DistillingDR, Ren2021RocketQAv2AJ,Hofsttter2021EfficientlyTA,Hofsttter2020ImprovingEN,Gao:2021:coCondenser} has shown that the performance of dense retrieval  models (i.e., \textit{the student models}) can be improved by distilling ranking knowledge from a more expensive re-ranking model (i.e., \textit{the teacher model}) that learns representations based on the interactions between query and document terms using cross-encoders~\cite{Nogueira2019PassageRW, Qiao2019UnderstandingTB}. 

In the knowledge distillation setting, an available teacher model assigns a distinct score to a query-document pair on which the supervision signal for optimizing the dense retrieval student model is based. Since the teacher can effectively score all pairs of queries and documents, we are not limited by the availability of labeled data, thereby providing us with greater flexibility. In this paper, we take advantage of this flexibility and introduce a generic \textit{curriculum learning} framework for training dense retrieval models via knowledge distillation.
The core idea of the curriculum learning (CL) is to provide a systematic approach to decompose the complex knowledge and design a curriculum for learning concepts from simple to hard~\cite{weng2020curriculum, Elman1993LearningAD,Krueger2009FlexibleSH}. 
Motivated by curriculum learning's ability to find better local optima~\cite{Bengio2009CurriculumL}, we propose a framework called \framework that introduces an iterative optimization process in which the difficulty level of the training data produced using the teacher model, as made available to the student, increases at every iteration. 
Through this \framework process, we first demand the dense retrieval student model to recover coarse-grained distinctions between the documents exampled by the teacher model and then progressively move towards recovering finer-grained ordering of documents.
For robust iterative optimization of the dense retrieval models, we adapt the listwise loss function of LambdaRank \cite{Burges2010FromRT} to our knowledge distillation setting. Therefore, our loss function only focuses on the order of documents produced by the teacher model, and not the exact document scores. 

In our experiments, we apply a simple implementation of the proposed optimization framework to two state-of-the-art dense retrieval models. First, we enhance TAS-B \cite{Hofsttter2021EfficientlyTA}, a model that uses a single representation vector for each query and document. Second, we repeat our experiments with the ColBERTv2 model \cite{Santhanam2021ColBERTv2EA}, a recent dense retrieval model that uses multiple representations per query and document. Our experiments on three public passage retrieval benchmarks demonstrate the effectiveness of the proposed framework. To improve the reproducibility of our models, we release the source code and the parameters of our models for research purposes.\footnote{\url{https://github.com/HansiZeng/CL-DRD}}

\section{Methodology}

\subsection{Background}
% \subsection{Overview}

\subsubsection{Dense Retrieval}
This paper focuses on the task of retrieving items based on high-dimensional dense learned representations for queries and documents, which is often called \emph{dense retrieval}. The query and document representations in dense retrieval models are often obtained using large-scale pre-trained language models, such as BERT~\cite{Devlin2019BERTPO}, fine-tuned for the downstream retrieval task~\cite{Karpukhin2020DensePR,Luan2021SparseDA,Xiong2021ApproximateNN}. Dense retrieval models can be seen as a category of vector space models \cite{Salton1975AVS}. Dense retrieval models compute the relevance score for a query $q$ and a document $d$ as follows:
$$\text{score}(q, d) = \text{sim}(E_\psi (q), E_\phi(d))$$
where $E_\psi(\cdot)$ and $E_\phi(\cdot)$ are the query encoders parameterized by $\psi$ and the document encoder parameterized by $\phi$, respectively. The encoders produce a dense representation of the given input. They often share the same output dimensionality: $|E_\psi (q)| = |E_\phi(d)|$. The similarity function `$\text{sim}$' is often implemented using the inner product. For efficient retrieval, dense retrieval models often employ approximate nearest neighbor (ANN) search algorithms.

\subsubsection{Knowledge Distillation}
For optimizing dense retrieval models, a ranking loss function is employed. The loss function is often based on either pointwise, pairwise, or listwise modeling, similar to learning-to-rank models. Since dense retrieval models often consist of millions or billions of trainable parameters, existing datasets are not often sufficient for training them. Recent work~\cite{Ren2021RocketQAv2AJ,Hofsttter2021EfficientlyTA,Lin2020DistillingDR} has successfully employed knowledge distillation (KD) for training dense retrieval models, where a teacher model produces training data for training a student model (i.e., the dense retrieval model). Teacher models are often more complex, with higher capacity and lower efficiency. A common approach is to use a cross encoding neural ranking model as the teacher model. In cross encoding models, both the given query and document are encoded jointly, leading to a superior result compared to dual encoders because of capturing more interaction information between the given query and document \cite{Nogueira2019PassageRW, Qiao2019UnderstandingTB}. We also use a cross encoding model in our experiments as the teacher. In more detail, we use a BERT model \cite{Nogueira2019PassageRW} that takes \texttt{[CLS] query tokens [SEP] document tokens [SEP]} as input and the relevance score is obtained by the linear transformation of the \texttt{[CLS]} representation.

\subsection{The \framework Framework}
Learning-to-rank models, including neural ranking models, are often trained based on the relevance judgment information. For instance, in pairwise models, a relevant document is paired with a sampled non-relevant document to form a training instance. However, in knowledge distillation, the training instances are produced based on the teacher model's output. This gives us substantial flexibility in producing the training data and we are not limited to the relevant documents that appeared in the relevance judgment file.

We take advantage of this flexibility and introduce the \framework framework. It combines the ideas of curriculum learning and knowledge distillation. \textbf{The intuition behind \framework is to introduce an iterative training process in which the difficulty of training data in each iteration increases.} This iterative optimization process is introduced in Algorithm~\ref{alg:framework}. In this algorithm, the parameter $\delta$ controls the difficulty of training sets generated by the `$\text{RankingDataGeneration}$' function. It starts with the difficulty level of 1 and its difficulty increases over time, until a stopping criterion is met. This stopping criterion can be based on the ranking performance on a held-out validation set, or can be based on a constant number of iterations. 

\framework is a generic framework and can be implemented in many different ways. For example, the difficulty of training sets can be modeled based on different assumptions. In this paper, we provide details of one of these implementations. In the following, we describe how we generate training data at each iteration (Algorithm~\ref{alg:framework}; line 6) and how we optimize the student model (Algorithm~\ref{alg:framework}; line 7). The other experimental details such as initialization (line 3) and the stopping criterion (line 10) are reported in Section~\ref{sec:exp}.

\begin{algorithm}[t]
\caption{The Iterative Optimization Process in \framework.}
\label{alg:framework}
\begin{algorithmic}[1]
\State \textbf{Input} (a) training query set $Q$; (b) document collection $C$; (c) \textit{optional} relevance judgment set $R$; (d) teacher model $\widehat{M}$.
\State \textbf{Output} dense retrieval model $M_\theta$. 
\State \textbf{Initialize} $\theta$.
\State $\delta \leftarrow 1$ \Comment{$\delta$ denotes the difficulty level in CL data}
\Repeat
\State $D_\delta \leftarrow \text{RankingDataGeneration}(\delta; \widehat{M}, Q, C, R)$
\State $\theta^* \leftarrow \arg \min_\theta \mathcal{L}_{KD}(M_\theta, D_\delta)$
%\State $\theta \leftarrow \theta^*$
\State $\delta \leftarrow \delta + 1$
\Until{stopping criterion is met}\\
\Return $M_{\theta^*}$
\end{algorithmic}
\end{algorithm}

\subsubsection{Generating Training Data in Each Iteration for Curriculum Learning}
\label{sec:CL}
\begin{figure}[t!]
    \centering
    \includegraphics[width=\linewidth]{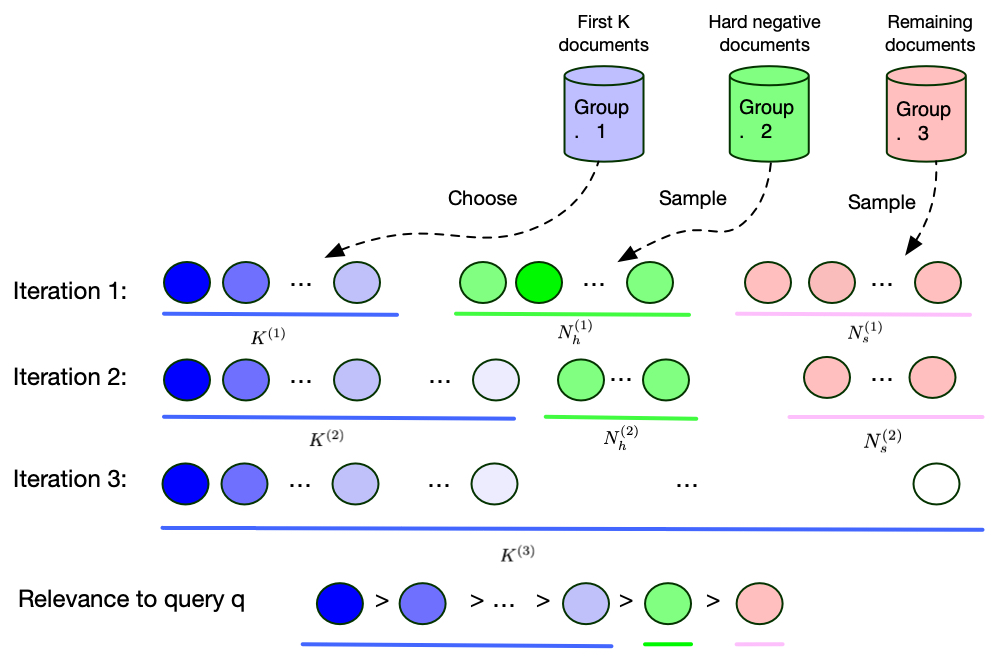}
    \caption{The data creation process in each iteration of curriculum learning based on knowledge distillation.}
    \label{fig:1}
    % \vspace{-1em}
\end{figure}

In typical curriculum learning, training instances are sorted or weighted according to their difficulty level and are fed to the model for optimization~\cite{Bengio2009CurriculumL, Matiisen2020TeacherStudentCL, MacAvaney2020TrainingCF,Penha:2020:CL}. Here, we use the same high-level idea, but with a substantially different approach. The `RankingDataGeneration' function in Algorithm~\ref{alg:framework} is supposed to produce more difficult ranking training data as $\delta$ increases. 
There are numerous ways to generate training data using knowledge distillation with different difficulty levels. Without loss of generality, this section describes the approach we choose for our experiments. In our approach, the training query set $Q$ remains the same for all iterations.\footnote{Iteration refers to the curriculum learning iterations in Algorithm~\ref{alg:framework}. Each iteration consists of many batches of training data.} For each training query $q \in Q$, we take the top $200$ documents returned by the student dense retrieval model and re-rank them using the teacher model, and then create three groups of documents: (1) the pseudo-relevant group: first $K$ documents returned by the teacher model, (2) the `hard negative' group: the next $K'$ documents in the ranked list returned by the teacher model, and (3) the remaining $K''$ documents in the ranked list produced by the teacher model. For each query, we keep the number of documents used for the optimization process constant and equal to $L$. Therefore, for training the student model in every iteration, we select $K$ documents from Group 1 (called $S_q^{(1)}$), we randomly sample $N_h$ documents from Group 2 (called $S_q^{(2)}$), and we finally randomly sampled $N_s$ documents from Group 3 (called $S_q^{(3)}$), where $L = K + N_h + N_s$. In the first iteration, we start with a relatively small value of $K$ and increase this value in subsequent iterations. Naturally, this leads to smaller $N_h+N_s$. Figure \ref{fig:1} provides an illustration of the process.

Therefore, for each query, we produce a ranked list of $L$ documents with the following pseudo-labels:
\begin{equation}
 y^t_q (d) = 
 \begin{cases}
    %\exp{(s^t_{qd})} & \quad \text{iff~} d \in S_q^{(1)} \\    
    \frac{1}{r^t_{qd}} & \quad \text{iff~} d \in S_q^{(1)} \\ 
    0 & \quad  \text{iff~} d \in S_q^{(2)} \\
    -1 & \quad  \text{iff~} d \in S_q^{(3)}
 \end{cases}
\end{equation}
where $r^t_{q,d}$ is the ranking position of the document $d$ given the query $q$ in the teacher ranked list.
%where $s^t_{qd}$ denotes the relevance score that the teacher assigns to the document $d$ and query $q$.
As we present in Section~\ref{sec:KD}, we use a loss function that only considers the order of document and the exact pseudo-label values do not impact the loss value.
We use a listwise loss function that does not rely on the exact document scores produced by the teacher model. 
To better understand our reasons for using such a pseudo-labeling strategy, let us categorize every document pairs with different pseudo-labels into four document pair types:
\begin{itemize}
    \item \textit{Type 1 pairs}: any two distinct documents from $S_q^{(1)}$. The document ranked higher by the teacher model is considered more relevant. This results in $\frac{K(K-1)}{2}$ document pairs for training.
    \item \textit{Type 2 pairs}: any document pair $d \in S_q^{(1)}$ and $d' \in S_q^{(2)}$. The document $d$ is considered more relevant. This results in $KN_h$ document pairs for training.
    \item \textit{Type 3 pairs}: any document pair $d \in S_q^{(1)}$ and $d' \in S_q^{(3)}$. The document $d$ is considered more relevant. This results in $KN_s$ document pairs for training.
    \item \textit{Type 4 pairs}: any document pair $d \in S_q^{(2)}$ and $d' \in S_q^{(3)}$. The document $d$ is considered more relevant. This results in $N_hN_s$ document pairs for training.
\end{itemize}

% In each iteration, we keep the number of passages in the passage pool the same, i.e., $|\mathcal{P}_q| = L$. There are $K$ of passages selected from the Top-$K$ passages, $N_h$ passages are sampled from the most-hard negative group and $N_s$ passages are sampled from the semi-hard negative group, and $L = K + N_h + N_s$.  In subsequent learning stages, we increase the value of $K$, which naturally leads to $N_h$ and $N_s$ decreasing. 

Learning from all $\textit{Type 2, 3, 4}$ pairs enforces the dense retrieval student model $M_\theta$ to distinguish documents from different groups, which is expected to be easier than than distinguishing document pairs in $\textit{Type 1}$. The reason is that Type 1 pairs enforce the student model to learn the exact ordering of documents provided by the teacher model. 

In our experiments, we consider three iterations of curriculum learning. From iteration 1 to 3, the value $K$ increases from $5$ to $10$ to $30$, respectively. In each iteration, the student model is trained for a fixed number of epochs. Since the number of pairs in Type 1 increases from $10$ to $45$ and to $435$ and Type 1 is expected to contain the most difficult document pairs, \textbf{the difficulty level of training data in each iteration is expected to increase.} Hence, in our curriculum learning based knowledge distillation algorithm, we progressively require that the student model to concentrate on more fine-grained differences in the output provided by the teacher model. %The difficulty level of the stage is quantified by the number of re-ranker labeled \textit{Type 1 pairs} that the training objective for the retriever is based upon. 

\subsubsection{Optimizing the Student Model through Knowledge Distillation}
\label{sec:KD}
Inspired by LambdaRank~\cite{Burges2010FromRT}, we use the following listwise loss function for training our student model using knowledge distillation at each iteration:
\begin{align}
   % \mathcal{L}_{KD}&(M_\theta, D) = \\
   % &\sum_{(q, S_q) \in D} \sum_{d, d' \in S_q} y^t_{q}(d, d') w(d,d') \log \sigma(M_\theta(q, d) - M_\theta(q, d')) \nonumber
    %
%     \displaystyle\sum_{q \in Q}\displaystyle\sum_{p, p' \in \mathcal{P}_q} \mathbbm{1}\{y^t_q(p) > y^t_q(p')\} w^s(p,p') \times \\  & \qquad \qquad \qquad \log \big(1 + e^{-(s_q(p) - s_q(p'))}\big) \numberthis \label{eq:kd-loss}\\  
%   w^s(p,p') &=  |\frac{1}{\pi^s_q(p)} - \frac{1}{\pi^s_q(p')}|  \\
%     & \text{where} \ \  \pi^s_q = \text{sort}(s_q)  
 \mathcal{L}_{KD}&(M_\theta, D_\delta) = \\
    &\sum_{(q, S_q) \in D_\delta} \sum_{d, d' \in S_q} y^t_{q}(d, d') w(d,d') \log ( 1+ e^{M_{\theta}(q, d') - M_{\theta}(q, d)}) \nonumber
\end{align}
where $S_q = S_q^{(1)} \cup S_q^{(2)} \cup S_q^{(3)}$ denotes all the documents selected via the pseudo-labeling approach presented in Section~\ref{sec:CL}, $y^t_{q}(d, d') = \mathbbm{1}\{y^t_q(d) > y^t_q(d')\}$. The function $w(d,d')$ is equal to $|\frac{1}{\pi_q(d)} - \frac{1}{\pi_q(d')}|$, where $\pi_q(d)$ denotes the rank of document $d$ in the result list produced by the student dense retrieval model $M_\theta$.

\section{Experiments}
\label{sec:exp}

\begin{table*}[t]
     \centering
     \caption{The performance comparison with state-of-the-art baselines. We use the two-tailed paired t-test with $p < 0.05$. The superscripts refer to significant improvements compared to all sparse retrieval models ($*$), ANCE and ADORE ($\dag$), TCT-ColBERT and Margin-MSE ($\ddag$), TAS-B ($\S$), and ColBERTv2 ($\P$). ``-'' denotes the results that are not applicable or available.}
    %  \vspace{-1em}
    % \vspace{-0.3cm}
     \setlength\tabcolsep{4pt}
    \begin{tabular}{llll!{\color{lightgray}\vrule}ll!{\color{lightgray}\vrule}ll!{\color{lightgray}\vrule}ll!}
    \toprule
    \multirow{2}{*}{\textbf{Model}} &
    \multirow{2}{*}{\textbf{KD}} &
    \multirow{2}{*}{\textbf{Encoder}} &
    \multirow{2}{*}{\textbf{\#params}} & 
    \multicolumn{2}{c!{\color{lightgray}\vrule}}{\textbf{MS MARCO DEV}} & 
       \multicolumn{2}{c!{\color{lightgray}\vrule}}{\textbf{TREC-DL'19}}&
       \multicolumn{2}{c}{\textbf{TREC-DL'20}}\\
       
    & &  &   &  MRR@10  & MAP@1k  & nDCG@10 & MAP@1k & nDCG@10 & MAP@1k \\
    
     \midrule
       \multicolumn{4}{l}{\textbf{Sparse Retrieval}} \\
        BM25 \cite{Robertson1995OkapiBM25} & -  & - & -  & .187 & .196 & .497 & .290 & .487 & .288 \\ 
       DeepCT \cite{Dai2019ContextAwareST} & -  &  - & -  & .243 & .250 & .550 & .341 & .556 & .343 \\ 
        docT5query \cite{Nogueira2019FromDT}  &  - &  - & -  & .272 & .281 & .642 & .403 & .619 & .407 \\
        %\arrayrulecolor{lightgray}
       \midrule
       \multicolumn{3}{l}{\textbf{Multi-Vector Dense Retrieval}} \\
     
       ColBERT \cite{Khattab2020ColBERTEA} & \xmark  &  BERT-Base & 110M & .360 & - & - & - & - & - \\
        ColBERTv2 \cite{Santhanam2021ColBERTv2EA} & \checkmark  & BERT-Base & 110M & \textbf{.397} & - & - & - & - & - \\ 
         ColBERTv2 \cite{Santhanam2021ColBERTv2EA} & \checkmark & DistilBERT & 66M & .384 & .389 & \textbf{.733} & .464 & .712 & .473 \\ 
        %  \multicolumn{3}{l}{\textbf{Ours}} \\
         ColBERTv2 + \framework (Ours) &  \checkmark & DistilBERT & 66M & .394$^{* \dag \ddag \S}$ & \textbf{.398}$^{* \dag \ddag \S \P}$ & .727$^{* \dag \ddag \S}$ & \textbf{.472}$^{* \dag \ddag \S \P}$ & \textbf{.717}$^{* \dag \ddag \S}$ & \textbf{.487}$^{* \dag \ddag \S \P}$ \\ 
         
       \midrule
      
       \multicolumn{3}{l}{\textbf{Single-Vector Dense Retrieval}} \\
        ANCE \cite{Xiong2021ApproximateNN}  & \xmark &  BERT-Base & 110M & .330 & .336 & .648 & .371 & .646 & .408 \\
        ADORE \cite{Zhan2021OptimizingDR} & \xmark   & BERT-Base & 110M & .347  & .352 & .683 & .419 & .666 & .442 \\ 
         RocketQA \cite{Qu2021RocketQAAO} & \checkmark &   ERNIE-Base & 110M  & .370 & - & - & - & - & -\\ 
        TCT-ColBERT \cite{Lin2020DistillingDR} & \checkmark  &  BERT-Base & 110M  & .335 & .342 & .670 & .391 & .668 & .430 \\ 
        Margin-MSE \cite{Hofsttter2020ImprovingEN} &  \checkmark  &  DistilBERT  & 66M & .325 & .331 & .699 & .405 & .645 & .416 \\
        TAS-B \cite{Hofsttter2021EfficientlyTA} & \checkmark  &  DistilBERT & 66M  &  .344  & .351 & .717 & .447 & .685 & .455 \\ 
        
        %\arrayrulecolor{lightgray}
        %\midrule
        %   \multicolumn{3}{l}{\textbf{Ours}}  \\
       
       TAS-B + \framework (Ours) & \checkmark &   DistilBERT & 66M  & \textbf{.382}$^{* \dag \ddag \S}$ & \textbf{.386}$^{* \dag \ddag \S}$  & \textbf{.725}$^{* \dag \ddag \S}$ & \textbf{.453}$^{* \dag \ddag }$ & \textbf{.687}$^{* \dag \ddag}$ & \textbf{.465}$^{* \dag \ddag \S}$ \\
       
       \arrayrulecolor{black}
       \bottomrule
    \end{tabular}
    \label{performance comparison}
    % \vspace{-0.3cm}
\end{table*}

\noindent \textbf{\underline{Datasets}}: We trained our model in the MS MARCO passage retrieval dataset \cite{Campos2016MSMA} which contains approximated $8.8$M passages and $503$K training queries with shallow annotations ($\approx$1.1 relevant passages per query on average). For the model evaluation, we use three datasets: (1) MS MARCO-Dev that contains $6980$ labeled queries, (2) TREC-DL'19: the passage retrieval dataset used in the 2019 edition of TREC Deep Learning Track \cite{trec2019overview} with 43 queries, and (3) TREC-DL'20: the passage retrieval dataset of TREC Deep Learning Track 2020 \cite{trec2020overview} with 54 queries. For evaluation, we report MAP@1000 for all three datasets, as well as the official metrics MRR@10 for MS MARCO and nDCG@10 for TREC-DL'19 and TREC-DL'20.

\medskip

\noindent \textbf{\underline{Experiment Settings}}: For the single-vector dense retrieval model, we use the DistilBERT \cite{Sanh2019DistilBERTAD} with the pre-trained checkpoint made available from TAS-B \cite{Hofsttter2021EfficientlyTA} as the initialization. For the multi-vector dense retrieval model, we also use the  DistilBERT \cite{Sanh2019DistilBERTAD} as the backbone. As the re-ranking teacher model, we use the MiniLM cross-encoder that is publicly available on HuggingFace.\footnote{\url{https://huggingface.co/cross-encoder/ms-marco-MiniLM-L-6-v2}}
We use the Adam optimizer~\cite{Kingma2015AdamAM} with  linear scheduling with the warmup of $4000$ steps and initial learning rate $[7e^{-6}, 3e^{-6}, 3e^{-6}]$ for the three CL iterations. We set the batch size to $8$ and the maximum length for queries and passages to $30$ and $256$ tokens, respectively. For the three iterations, the sizes of group $1$: $[5,10,30]$, group $2$: $[45, 40, 20]$, group $3$: $[150,150,150]$. The number of sampled documents for each query: $L=30$, and $K = [5,10,30]$, $N_h = [12,10,0]$, $N_s = [13,10,0]$ for the three iterations. 

% Then we use the re-ranker to rerank\footnote{We use a MiniLM cross-encoder trained with distillation from \url{https://huggingface.co/cross-encoder/ms-marco-MiniLM-L-6-v2}} the Top -200 passages retrieved by the retriever for each query. The training epochs for the $3$ learning stages are $[4, 2, 2]$, and we use the Adam\cite{Kingma2015AdamAM} with  linear scheduling with the warmup of $[4000, 4000, 4000]$ steps and initial learning rate $[7e^{-6}, 3e^{-6}, 3e^{-6}]$. We set the batch size as $8$ and the max lengths for input queries and passages are $30$, $256$.

\begin{comment}
\noindent \textbf{\underline{Baselines}}: We compare our models to three categories of baselines. (1) Retrieval models with sparse representations: we adopt BM25~\cite{Robertson1995OkapiBM25}, DeepCT~\cite{Dai2019ContextAwareST} and docT5query~\cite{Nogueira2019FromDT}. (2) State-of-the-art dense retrieval models with a single representation vector for each query and document: we adopt ANCE \cite{Xiong2021ApproximateNN}, ADORE \cite{Zhan2021OptimizingDR}, RocketQA \cite{Qu2021RocketQAAO}, TCT-ColBERT \cite{Lin2020DistillingDR}, Margin-MSE \cite{Hofsttter2020ImprovingEN} and TAS-B \cite{Hofsttter2021EfficientlyTA}. (3) State-of-the-art dense retrieval models with multiple representations for each query and document: we adopt ColBERT~\cite{Khattab2020ColBERTEA} and ColBERTv2~\cite{Santhanam2021ColBERTv2EA}. Note that we face difficulties reproducing the ColBERTv2 experiments using BERT-Base, thus we additionally provide the results for our own implementation of the model using DistilBERT.
\end{comment}

\medskip

\noindent \textbf{\underline{The \framework models}}: \framework is a generic optimization framework for dense retrieval models that can be applied to any dense retrieval model. In our experiments, we trained two different dense retrieval models using \framework: (1) \textbf{TAS-B + \framework}: TAS-B is the best performing dense retrieval baseline that uses a \textit{single vector} representation for each query and document. In TAS-B + \framework we apply our framework to the TAS-B model. (2) \textbf{ColBERTv2 + \framework}: Similarly, we choose ColBERTv2 the best performing dense retrieval model that uses \textit{multiple vectors} per query and document. Note that this model uses our own implementation of ColBERTv2 that uses DistilBERT. We compare our models with several state-of-the-art baselines shown in the Table~\ref{performance comparison}.

% To compare with lexical matching baselines, we adopt the Anserini's implementation of BM25~\cite{Robertson1995OkapiBM25}, DeepCT~\cite{Dai2019ContextAwareST} and docT5query~\cite{Nogueira2019FromDT}. To compare with state-of-art dense retrieval models without  knowledge distillation, we use ANCE \cite{Xiong2021ApproximateNN}, ADORE \cite{Zhan2021OptimizingDR}. To compare with dense retrieval models with knowledge distillation, we select RocketQA \cite{Qu2021RocketQAAO}, .

\medskip

\noindent \textbf{\underline{Results and Discussion}}:  The retrieval results obtained by our models and the baselines are reported in Table \ref{performance comparison}. 
According to the results, dense retrieval models generally outperform lexical matching models including the ones that use pre-trained language models for document expansion, such as docT5query \cite{Nogueira2019FromDT}. TAS-B + \framework outperforms all the dense retrieval baselines that use a single vector for representing each query and document. This improvement is consistent across all three datasets, and is statistically significant in most cases. Note that some of these baseline models use significantly larger models compared to ours. The number of parameters for each model is mentioned in Table \ref{performance comparison}. Since TAS-B + \framework uses TAS-B \cite{Hofsttter2021EfficientlyTA} as its parameter initialization, we can have a direct comparison with this baseline. Applying \framework to TAS-B leads to 11\% improvements on the MS MARCO Dev set in terms of MRR@10. 

\tablename~\ref{performance comparison} suggests that \framework can also improve dense retrieval models with multiple vector representations. When we apply \framework to ColBERTv2 (the current state-of-the-art in dense retrieval with DistilBERT), we obtain improvements in terms of all metrics on all collections except for nDCG@10 on TREC DL'19. The improvements are larger for recall-oriented metrics, e.g., MAP@1000, and they are statistically significant in majority of cases. This demonstrates that \framework is sufficiently flexible to be applied to different dense retrieval models and introduce significant improvements.

In general, ColBERTv2 + \framework performs better than TAS-B + \framework. It is worth noting that representing queries and documents with multiple vectors demands significantly higher memory requirements, indexing cost, and query processing cost.

% query time and memory for searching and indexing, our model can achieve very close performance, but use less resources. The best model in the category by far, ColBERTv2 \cite{Santhanam2021ColBERTv2EA}, requires $30$ GB memory for indexing all $8.8$ M passages of MS MARCO dataset, which is x$2.3$ larger than the memory needed ($13$ GB ) of our model, but only achieves $0.5\%$ improvement over MRR@10. (3) In contrast to other knowledge distillation models, our model does not requires access to the re-ranker predicted scores, which might imply that with the proper learning technique--curriculum learning, the retriever can further improve its ranking effectiveness by imitating the re-ranker's pairwise preference. 

\begin{figure}
    %\centering
    \includegraphics[scale=.165]{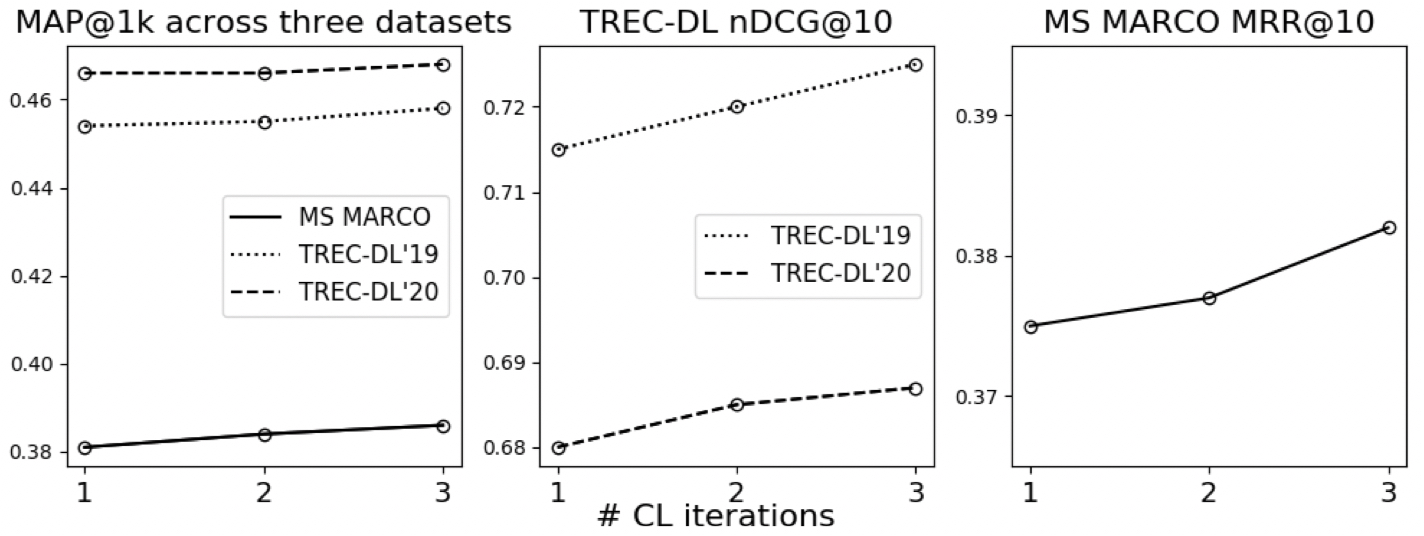}
    \caption{The results obtained by TAS-B + \framework at different iterations of curriculum learning.}
    \label{fig:ablation}
    % \vspace{-2em}
\end{figure}

\medskip

\noindent \textbf{\underline{Ablation Study}}: 
We conduct a few experiments for evaluating the effectiveness of curriculum learning. For the sake of space, we solely focus on the TAS-B + \framework model. Figure~\ref{fig:ablation} demonstrates the performance of this model after each curriculum learning iteration on all three datasets. As shown in the figure, the performance generally improves at each iteration. This demonstrates that the proposed CL approach is effective for training ranking models. To make sure that this improvement is not just due to the number of epochs or the size of training sets, we additionally train our model on a reverse curriculum learning setup. In this experiment, we followed the same procedure with the same number of iterations, but we start from the last iteration used in our CL approach. We observe that our model still outperforms this reverse CL approach. For example, this approach achieves an nDCG@10 of 0.715 and 0.683 on TREC DL'19 and TREC DL'20, respectively. It also achieves an MRR@10 of 0.378 on MS MARCO Dev set, which is significantly lower than the results obtained by our method.

\section{Conclusions and Future Work}
We introduced \framework, a generic framework for optimizing dense retrieval models through knowledge distillation. Inspired by curriculum learning, \framework follows an iterative process in which supervision of increasing levels of difficulty are derived from the teacher model's output. We provided a simple implementation of this framework and demonstrated its effectiveness on three public passage retrieval benchmarks. 

In the future, we intend to explore more sophisticated solutions for controlling the difficulty of each iteration in \framework. We are also interested in developing machine learning models that can select informative training instances based on the teacher's performance. 

\section{Acknowledgments}
This work was supported in part by the Center for Intelligent Information Retrieval, and in part by gift funding from Adobe Research. Any opinions, findings and conclusions or recommendations expressed in this material are those of the authors and do not necessarily reflect those of the sponsor.

\begin{comment}
\begin{table}[h]
   \caption{Ablation study for curriculum learning.}
    \begin{tabular}{c!{\color{lightgray}\vrule}c!{\color{lightgray}\vrule}c!{\color{lightgray}\vrule}c!}
        
        \toprule

       \multirow{2}{*}{\textbf{Training}} & \multicolumn{1}{c!{\color{lightgray}\vrule}}{\textbf{MS MARCO DEV}} & 
       \multicolumn{1}{c!{\color{lightgray}\vrule}}{\textbf{TREC-DL'19}}&
       \multicolumn{1}{c}{\textbf{TREC-DL'20}}\\
        & MRR@10  &  nDCG@10  &  nDCG@10 \\
       
       \midrule
       \multicolumn{2}{l!}{\textbf{Partial-curriculum}} \\
       \arrayrulecolor{lightgray}
       
       T5 & .375 &  .715 &  .680 \\
       T10 & .378  & .714  & .679  \\
       T30 & .377  & .710  & .670  \\ 
       
       T5 $\xrightarrow[]{}$T10 & .377  & .720  & .685  \\ 
       T10 $\xrightarrow[]{}$T30 & .380  & .717 & .680  \\ 
       
       \midrule
       \multicolumn{2}{l!}{\textbf{Anti-curriculum}} \\ 
       T30$\xrightarrow[]{}$T10$\xrightarrow[]{}$T5 & .378 &  .715   & .683  \\ 
       \midrule
       T5 $\xrightarrow[]{}$ T10 $\xrightarrow[]{}$T30 & \textbf{.382} &   \textbf{.725} &   \textbf{.687}  \\ 
       
        \arrayrulecolor{black}
        \bottomrule
    \end{tabular}
    \label{ablation study}
\end{table}

\newpage 
\end{comment}
% \begin{figure}
%     %\centering
%     \caption{Ablation study for curriculum learning w/ TAS-B}
%     \includegraphics[width=0.45\textwidth]{figures/suite.png}
%     \label{fig:my_label}
% \end{figure}

\bibliographystyle{ACM-Reference-Format}
\bibliography{sigproc}
\end{document}